\pdfoutput=1
\documentclass[sigconf,screen,hyphens]{acmart}

\AtBeginDocument{%
  }

\setcopyright{rightsretained}
\copyrightyear{2024}
\acmYear{2024}
\acmDOI{10.1145/3613904.3642703}

\acmConference[CHI '24]{Proceedings of the CHI Conference on Human Factors in Computing Systems}{May 11--16, 2024}{Honolulu, HI, USA}
\acmBooktitle{Proceedings of the CHI Conference on Human Factors in Computing Systems (CHI '24), May 11--16, 2024, Honolulu, HI, USA}
\acmPrice{15.00}
\acmISBN{979-8-4007-0330-0/24/05}

\usepackage{orcidlink}
\hypersetup{breaklinks=true}

\begin{document}

\title{The Illusion of Artificial Inclusion}

\author{William Agnew \orcidlink{0000-0002-1362-554X}}
\affiliation{%
  \institution{Carnegie Mellon University}
  \city{Pittsburgh}
  \state{PA}
  \country{USA}
}
\author{A. Stevie Bergman \orcidlink{0000-0002-4331-1357}}
\affiliation{%
  \institution{Google DeepMind}
  \city{New York}
  \state{NY}
  \country{USA}
}
\author{Jennifer Chien \orcidlink{0009-0009-8768-1761}}
\affiliation{%
  \institution{University of California San Diego}
  \city{San Diego}
  \state{CA}
  \country{USA}
}
\author{Mark D\'{i}az \orcidlink{0000-0003-0167-9839}}
\affiliation{%
  \institution{Google Research}
  \city{New York}
  \state{NY}
  \country{USA}
}
\author{Seliem El-Sayed \orcidlink{0000-0003-4819-1136}}
\affiliation{%
  \institution{Google DeepMind}
  \city{London}
  \country{UK}
}
\author{Jaylen Pittman \orcidlink{0000-0002-5176-5922}}
\affiliation{%
  \institution{Stanford University}
  \city{Stanford}
  \state{CA}
  \country{USA}
}
\author{Shakir Mohamed \orcidlink{0000-0002-1184-5776}}
\affiliation{%
  \institution{Google DeepMind}
  \city{London}
  \country{UK}
}
\author{Kevin R. McKee \orcidlink{0000-0002-4412-1686}}
\affiliation{%
  \institution{Google DeepMind}
  \city{London}
  \country{UK}
}

\renewcommand{\shortauthors}{W. Agnew, A.S. Bergman, J. Chien, M. D\'{i}az, S. El-Sayed, J. Pittman, S. Mohamed, \& K.R. McKee}

\begin{abstract}
  Human participants play a central role in the development of modern artificial intelligence (AI) technology, in psychological science, and in user research. Recent advances in generative AI have attracted growing interest to the possibility of replacing human participants in these domains with AI surrogates. We survey several such ``substitution proposals'' to better understand the arguments for and against substituting human participants with modern generative AI. {Our scoping review} indicates that the recent wave of these proposals is motivated by goals such as reducing the costs of research and development work and increasing the diversity of collected data. However, these proposals ignore and ultimately conflict with foundational values of work with human participants: representation, inclusion, and understanding. This paper critically examines the principles and goals underlying human participation to help chart out paths for future work that truly centers and empowers participants.
\end{abstract}



\maketitle

\section{Introduction}

Participation is a foundational element of the social-behavioral sciences and in the design of new technology. In psychology, user research, human-computer interaction (HCI), and other related fields, research participants offer a window into human cognition and decision making. In the development of new technologies, human participants ground the design process in real-life needs, perspectives, and experiences.

The past year has seen a growing number of proposals recommending the replacement of human participants in technology development and scientific research with large language models (LLMs), a new class of artificial intelligence (AI) systems. These proposals include at least {thirteen} technical reports and peer-reviewed research articles, jointly sharing {over one thousand} citations at the time of this manuscript's submission~\citep{aher2023using, argyle2023out, bai2022constitutional, byun2023dispensing, chiang2023can, dillion2023can, gerosa2023can, gilardi2023chatgpt, hamalainen2023evaluating, heyman2023impact, horton2023large, park2022social, wang2021want}, in addition to several commercial products~\citep{opinioai, syntheticusers, userpersona}.

The momentum building behind these proposals echoes broader social reactions to modern AI systems: LLMs have shocked and fascinated both AI researchers and the general public with their ability to produce fluent, human-like text in a range of domains~\citep{roose2023conversation, thompson2022breakthroughs, tiku2022google}. In this paper, we seek to understand and contextualize the arguments for replacing human participants with LLMs and other generative AI systems. Our {thematic content analysis} suggests that these proposals are motivated by aims such as reducing the costs of scientific research, protecting participants from potential harms, and augmenting diversity in data collection. However, a critical examination of the social structures and epistemic frameworks surrounding human participation reveals fundamental conflicts between these substitution proposals and the values underpinning research and development. In particular, substitution undermines three values: the \textit{representation} of participants' interests; participants' \textit{inclusion} and empowerment in the development process; and the \textit{understanding} that researchers otherwise develop through intersubjective engagement with participants. AI-generated substitutes cannot meaningfully support technological development and scientific research in the absence of a deeper reckoning with their relationship to representation, inclusion, and understanding. While current proposals undermine these values, there may be space for scientific and engineering communities to re-imagine ways for generative AI to support human participation. Overall, our analysis underscores the need for researchers and developers to take action at every stage of their work to center and empower participants and the communities to which they belong.

\section{Background}

{\subsection{Historic applications of artificial intelligence in social-behavioral research}}

{Advances in artificial intelligence closely intertwine with the histories of psychology and user research~\citep{grudin2009ai, grudin2022tool, newell1982intellectual}. As early as the 1980s, scientists envisioned a key role for AI in supporting psychology research (e.g.,~\citep{schmidt1978plan, van1988artificial}), as well as enabling technological development and advancing user experience (e.g.,~\citep{belkin1987knowledge, bing1986text}). Over the following decades, the increasing availability of machine learning---AI techniques that solve tasks by adapting to incoming data, in contrast to early rule-based approaches---opened up new possibilities for AI applications in user research and in neighboring social-behavioral fields. Early machine learning algorithms were relatively rudimentary, but nonetheless enabled the automation of various tasks in the research process. For example, new AI tools allowed scientists to quickly analyze sentiment in open-ended text data~\citep{pang2002thumbs} and to make fine-grained predictions of user preferences~\citep{heckerman2000dependency}.}

{Today, applications of AI pervade throughout user research and psychology. Many occurrences likely seem pedestrian, including common statistical techniques such as regression, factor analysis, and classification. The research community has also pushed a number of boundaries with new, creative applications of AI. For example, researchers have leveraged machine learning to automate the development of user personas~\citep{zhang2016data} and accelerate design iteration~\citep{buschek2020paper2wire}. In psychology, AI methods have helped to scale the coverage of traditional experimental tools~\citep{nicolas2021comprehensive} and explore novel solutions to persistent behavioral puzzles~\citep{mckee2023scaffolding}. Overall, recent innovations expand the focus of AI applications from predominantly supporting analysis to aiding the full discovery process~\citep{stige2023artificial}.}

\subsection{The emergence and capabilities of large language models}

Recent breakthroughs in AI research, most notably the emergence of large language models (LLMs; e.g.,~\citep{brown2020language, chowdhery2022palm, devlin2019bert}), have invigorated new interest in real-world applications of AI systems. At their core, LLMs are natural language processing (NLP) systems: a user can prompt an LLM with text, and the LLM will output text in response. Developing an AI system capable of proficiently conversing with humans has been a longstanding aspiration of AI researchers (e.g.,~\citep{turing1950computing}), and NLP systems have existed for decades, in forms including chatbots, text completion, and search engines. However, LLMs represent a substantial breakthrough in language capabilities for AI systems. LLMs can generate text that is fluent, grammatically correct, and surprisingly human-like, in response to a dizzying range of inputs (e.g.,~\citep{openai2023gpt4technical}). This flexibility enables them to perform remarkably well in applications including summarizing books~\citep{anthropic2023introducing}, generating computer code~\citep{schulman2022chatgpt}, and conversing with users for some specific tasks~\citep{thoppilan2022lamda}, though serious problems persist (e.g.,~\citep{ji2023survey}). Some contemporary models can process and generate data in multiple formats, including not just text, but also images and audio~\citep{alayrac2022flamingo}. These multimodal models are often grouped with language-only models in the broader category of ``generative AI''~\citep{stokel2023chatgpt}.

The capabilities of LLMs and generative AI have fascinated researchers, policymakers, and members of the public~\citep{novak2023viral, roose2023conversation, tiku2022google, vincent2022all}. LLMs have also courted controversy, particularly over claims that they ``understand'' natural language (e.g.,~\citep{manning2022human, y2022large}) and exhibit signs of general intelligence (e.g.,~\citep{bubeck2023sparks}). Critics argue that such claims of understanding and general intelligence generate overconfidence in model abilities, which in turn could deter efforts at oversight and encourage unsafe or inappropriate applications~\citep{bender2021dangers, shah2022situating, weidinger2022taxonomy}. Regardless of these grander debates, in their capability to produce fluent and human-like text for various tasks in English and other highly datafied languages, LLMs represent a large advance over prior approaches to NLP.

\subsection{Human participants in the context of AI development and social-behavioral research}

Humans interact with the AI development lifecycle at multiple points~\citep{nist2023artificial}, including by driving the development process itself (as scientists, engineers, and other development team members); contributing data to train models~\citep{hutchinson2021towards, paullada2021data, rolf2021representation, sambasivan2021seeing, sambasivan2021everyone}; evaluating models as annotators, auditors, and testers~\citep{bergman2022towards, mckee2023human, partnership2021responsible}; and reviewing products and outcomes as system stakeholders~\citep{fazelpour2022diversity}. Similarly, in psychology, user research, and other social-behavioral fields, humans play key roles across multiple stages of the research process, including in ideation, hypothesis generation, and study design~\citep{coan1979psychologists, jamieson2023reflexivity}; data provision and collection~\citep{danziger1994constructing}; and the interpretation of study results~\citep{ziman1991public}.

Across both of these domains, ``human participants'' are the class of people who provide information and feedback to development work and research projects, typically as distinguished from the developers and researchers themselves.\footnote{This distinction is stronger today than at various points in the past. Notably, early efforts in psychology and supervised machine learning often blurred the roles of researcher and participant. Before the advent of online participant platforms, for instance, NLP research teams frequently took responsibility for annotating training datasets themselves~\citep{shmueli2021beyond}. Similarly, in the earliest years of psychology research, scientists often served the roles of both investigator and participant~\citep{walsh2014researcher}.} In AI development, they are the individuals ``whose judgment and intelligence are widely employed [...] to train and validate models''~\citep{partnership2021responsible}, as well as members of communities who review outcomes as project stakeholders (e.g.,~\citep{zytko2022participatory}). In the behavioral and social sciences, human participants are people who provide information and data to researchers, particularly in systematic investigations intended to produce generalizable insights and knowledge~\citep{us2018code}. Traditionally, research participants interact with scientists through channels including laboratory studies and experiments, focus groups, interviews, and surveys. For a detailed historical examination of the role of human participants in social and behavioral research, see~\citep{danziger1994constructing}.

\section{Reviewing substitution proposals and their stated motivations}

Enthusiasm for the capabilities of LLMs has motivated a wave of real-world applications of AI, including systems designed to assist scientists and engineers. For instance, several recent models offer support to research scientists and AI developers through article summarization~\citep{anthropic2023model} and code generation~\citep{chen2021evaluating}. A nascent class of these proposals---emerging across academic papers, technical reports, and startup products---suggests using LLMs to take over the roles of human participants in research and development. The defining feature of these ``substitution proposals'' is the idea that LLMs can simulate or replicate the cognition, decision making, and attitudes of human participants in order to fulfill a role that participants normally play in a project.\footnote{Notably, this scope excludes social simulacra projects that study LLMs as social or intelligent agents in themselves (cf.~\citep{levy1992artificial}), rather than as a replacement for human participants in research and development. For instance, \citet{park2023generative} investigate the emergent social dynamics between LLM-based agents rather than testing any predictions about human behavior or replacing inputs to development.} For example, to empirically test the possibility of substitution in HCI, \citet{hamalainen2023evaluating} collected materials from a previous user study with human participants and reformulated them as appropriate inputs for an LLM. The authors passed these inputs to an LLM, parsed the model’s outputs as synthetic user responses, and then analyzed and interpreted those responses as they would from human participants.

{We conducted a scoping review to better understand the arguments motivating AI substitution, sampling and evaluating a set of prominent papers and sources that advance substitution proposals. Our goal was to survey the different arguments offered for substitution by identifying the motivations set out in the text of the proposals themselves. To conduct this review, we collected an initial purposive sample of proposals, then leveraged snowball sampling to expand to a final set for analysis~\citep{atkinson2001accessing}. Our initial sample included eight proposals~\citep{byun2023dispensing, dillion2023can, gilardi2023chatgpt, horton2023large, opinioai, syntheticusers, userpersona, wang2021want}, each of which gained visibility after publication in disciplinary venues that we monitor (e.g., \textit{Proceedings of the National Academy of Sciences} and \textit{Trends in Cognitive Sciences}) or through discussion on social media platforms (e.g.,~\citep{infoxiao2023trend, schock2023now}). We subsequently employed snowball sampling to expand our survey, iteratively identifying additional proposals cited by the sources in our sample (backward connections) and referring to sources in our sample (forward connections).
We focused the review on proposals with a demonstrated interest in testing or pursuing AI substitution, excluding sources that conceptually discuss substitution, but that do not empirically explore the idea.
Our final sample comprised sixteen sources, including thirteen technical reports and research articles, and three commercial products (Table~\ref{tab:substitution_proposals}).}

\begin{table*}[t]
    \centering
    \def\arraystretch{1.1}
    \begin{tabular}{|l|l|l|l|l|}
        \hline
        \multicolumn{1}{|c|}{\textbf{Proposal}} & \multicolumn{1}{|c|}{\textbf{Motivation(s) discussed}} & \multicolumn{1}{|c|}{\textbf{{Position}}} & \multicolumn{1}{|c|}{\textbf{Application domain}} & \multicolumn{1}{|c|}{\textbf{{Venue}}} \\ \hline
        \citet{aher2023using} & \begin{tabular}[c]{@{}l@{}}Increase speed,\\ reduce costs,\\ augment diversity,\\ protect participants\end{tabular} & {In favor} & Behavioral science & {\begin{tabular}[c]{@{}l@{}}\textit{Int. Conf. Mach.}\\ \textit{Learn.} (ICML)\end{tabular}} \\ \hline
        \citet{argyle2023out} & Augment diversity & {In favor} & Survey research & \textit{{Political Anal.}} \\ \hline
        \citet{bai2022constitutional} & \begin{tabular}[c]{@{}l@{}}Increase speed,\\ protect participants\end{tabular} & {In favor} & AI development & \textit{{arXiv*}} \\ \hline
        \citet{byun2023dispensing} & \begin{tabular}[c]{@{}l@{}}Increase speed,\\ augment diversity,\\ protect participants\end{tabular} & {Mixed} & \begin{tabular}[c]{@{}l@{}}Human-computer\\ interaction\end{tabular} & {\begin{tabular}[c]{@{}l@{}}\textit{Conf. Hum. Factors}\\ \textit{Comput. Syst.} (CHI)\end{tabular}} \\ \hline
        \citet{chiang2023can} & \begin{tabular}[c]{@{}l@{}}Increase speed,\\ reduce costs,\\ protect participants\end{tabular} & {In favor} & AI development & {\begin{tabular}[c]{@{}l@{}}\textit{Annu. Meet. Assoc.}\\ \textit{Comput. Linguist.} (ACL)\end{tabular}} \\ \hline
        \citet{dillion2023can} & Increase speed & {Mixed} & Psychological science & \textit{{Trends Cogn. Sci.}} \\ \hline
        {\citet{gerosa2023can}} & \begin{tabular}[c]{@{}l@{}}{Increase speed,}\\ {augment diversity,}\\ {protect participants}\end{tabular} & {Mixed} & {Software engineering} & \textit{{arXiv*}} \\ \hline
        \citet{gilardi2023chatgpt} & Reduce costs & {In favor} & AI development & \textit{{Proc. Natl. Acad. Sci.}} \\ \hline
        \citet{hamalainen2023evaluating} & \begin{tabular}[c]{@{}l@{}}Increase speed,\\ reduce costs,\\ augment diversity\end{tabular} & {Mixed} & \begin{tabular}[c]{@{}l@{}}Human-computer\\ interaction\end{tabular} & {\begin{tabular}[c]{@{}l@{}}\textit{Conf. Hum. Factors}\\ \textit{Comput. Syst.} (CHI)\end{tabular}} \\ \hline
        {\citet{heyman2023impact}} & {Increase speed} & {Critical} & {Psychological science} & \textit{{Behav. Res. Methods}} \\ \hline
        {\citet{horton2023large}} & \begin{tabular}[c]{@{}l@{}}{Increase speed,}\\ {reduce costs}\end{tabular} & {In favor} & {Experimental economics} & \textit{{Natl. Bur. Econ. Res.*}} \\ \hline
        \citet{opinioai} & \begin{tabular}[c]{@{}l@{}}Increase speed,\\ reduce costs,\\ augment diversity\end{tabular} & {In favor} & Marketing research & {Product website} \\ \hline
        \citet{park2022social} & \begin{tabular}[c]{@{}l@{}}Increase speed,\\ protect participants\end{tabular} & {Mixed} & \begin{tabular}[c]{@{}l@{}}Human-computer\\ interaction\end{tabular} & {\begin{tabular}[c]{@{}l@{}}\textit{Symp. User Interface}\\ \textit{Softw. Technol.} (UIST)\end{tabular}} \\ \hline
        \citet{syntheticusers} & \begin{tabular}[c]{@{}l@{}}Increase speed,\\ reduce costs,\\ augment diversity\end{tabular} & {In favor} & User research & {Product website} \\ \hline
        \citet{userpersona} & \begin{tabular}[c]{@{}l@{}}Increase speed,\\ augment diversity\end{tabular} & {In favor} & Marketing research & {Product website} \\ \hline
        \citet{wang2021want} & \begin{tabular}[c]{@{}l@{}}Increase speed,\\ reduce costs\end{tabular} & {In favor} & AI development & {\begin{tabular}[c]{@{}l@{}}\textit{Empir. Methods Nat.}\\ \textit{Lang. Process.} (EMNLP)\end{tabular}} \\ \hline
    \end{tabular}
    \caption{A summary of the surveyed proposals to substitute human participants in research and development with large language models (LLMs) and generative artificial intelligence (AI). {The ``Motivation(s) discussed'' column lists the potential goals of substitution explicitly discussed by each source. ``Position'' reflects whether the source overall voices support for simulating human participants, criticism for it, or a combination of optimism and concern. ``Application domain'' describes the field or area in which the source considers simulating participation. For papers and technical reports, ``Venue'' provides the ISO 4 abbreviation for the conference, journal, or repository where the proposal was shared, with common abbreviations in parentheses; asterisks indicate technical reports. For products, ``Venue'' simply states ``product website''.}}
    \label{tab:substitution_proposals}
\end{table*}

{These substitution proposals focus on human participation in AI development~\citep{chiang2023can, gilardi2023chatgpt} and in a range of scientific fields including user research~\citep{hamalainen2023evaluating, byun2023dispensing, syntheticusers}, social and cognitive psychology~\citep{dillion2023can}, and survey research~\citep{argyle2023out}.
While we do not claim that this survey represents an exhaustive review, the diversity of the surveyed sources helps provide a broad perspective on the idea of substituting generative AI for human participation. If substitution proposals continue to emerge over time, future research could test the representativeness of our survey themes through a systematic literature review.}

{\subsection{Evaluating manifest motivations for substitution}}

{We take an iterative, inductive, and semantic approach to code the themes in this sample of substitution proposals. Our protocol aims to identify themes through a process of repeated review and refinement, without imposing pre-existing concepts or categories (iterative and inductive~\citep{braun2006using}), focusing on the explicit contents of the text (semantic; also known as ``manifest content''~\citep{berelson1952content, potter1999rethinking}). We independently read the proposals, noting any quality concerns that might affect the depth or rigor of our analysis. We separately coded passages that communicate explicit motivations for substitution, then jointly discussed and organized these text data into themes. We returned to the text of the sources during these discussions to ensure alignment on our final thematic codes. We also discussed and coded the overall position that each source adopted on AI substitution: overall in favor (offering limited or no reservations about substitution), overall critical (indicating limited or no endorsement of substitution), or mixed (sharing meaningful support alongside meaningful critique or concern).}

{Table~\ref{tab:substitution_proposals} summarizes our overall findings. As a preliminary step for our analysis, we found the sources in our review to be generally clear and coherent. Our sample included, for example, ten research articles that indicated undergoing and passing academic peer review. They, along with the three non-peer reviewed technical reports and the three product websites, were well-written and easy to follow, communicating their arguments effectively. Most sources articulated a position somewhere between guarded excitement (``Practically speaking, LLMs may be most useful as participants when studying specific topics, when using specific tasks, at specific research stages, and when simulating specific samples''~\citep{dillion2023can}) and unbridled optimism (``These language models can be used prior to or in the absence of human data''~\citep{argyle2023out}). We coded 10 sources as overall in favor of substitution, five with a mixture of support and concern, and only one overall critical of the idea.}

{Every article communicated at least one potential motivation for simulating participants with LLMs. Across the entire sample, our thematic analysis identified four recurring motivations for replacing human participants with generative AI.}

\subsubsection{``Substitution increases the speed {and scale} of research and development''}

{Fourteen proposals---nearly every source in our sample---discuss the possibility of increasing the speed and scale of research or development as a motivation for replacing participants with LLMs. For instance, \citet{wang2021want} cite the ``time-consuming and labor-intensive'' nature of work with human participants as crucial context for preferring to employ state-of-the-art models, since they ``can label data non-stoppingly at a much faster speed than human labelers''. \citet{chiang2023can} similarly describe simulated participants as ``cheaper and faster than human evaluation'', and \citet{dillion2023can} admire how ``LLMs can rapidly answer hundreds of questions without fatigue''. \citet{bai2022constitutional} argue generative AI can ``scale supervision'' and ``reduce iteration time by obviating the need to collect new human feedback labels when altering the [feedback] objective''. \citet{userpersona} touts the capability of synthesized participants ``to unlock more time for high-value work'', and \citet{horton2023large} describes ``the advantages in terms of speed'' as ``enormous''. Pointing out that the majority of researchers who work with human participants are ``time-limited'', \citet{syntheticusers} offers pithy advice: ``simulate to accelerate''.}

{The proposals frequently conditioned these speed and scale arguments on claims about the accuracy of LLMs. Immediately after identifying ``high speed'' as a potential benefit of substitution, \citet{hamalainen2023evaluating} muse that ``data quality'' might prove to be a concomitant drawback. \citet{gerosa2023can} state that ``if [software engineers] could simulate [...] human behaviors with sufficient accuracy, the potential for scaling research efforts could be unprecedented''. Finally, \citet{aher2023using} propose that future ``increase[s] in accuracy'' with LLMs can help alleviate ``considerations regarding scale'' in behavioral research.}

\subsubsection{``Substitution lowers costs and reduces reliance on human labor''}

{Reducing cost was another prominent argument, referenced in eight of the proposals. \citet{gilardi2023chatgpt} estimate that the cost of LLM surrogates is on the order of ``thirty times cheaper than MTurk'', a standard online platform for recruiting human participants. Likewise, \citet{hamalainen2023evaluating} laud the possibility of ``low cost and high speed of LLM data generation'' in the context of user research. Four sources connect this point to a related motivation: reducing the need for manual labor. As argued by \citet{chiang2023can}, substitution alleviates development costs by reducing the need for humans in data collection. \citet{aher2023using} similarly note the intertwined benefits of limiting monetary costs and reducing reliance on effort by human participants. Interestingly, these considerations appear to primarily emerge in proposals for AI development, HCI, and commercial products.}

{Mirroring the discussion of speed and scale advantages, arguments about cost efficiency were frequently paired with assertions concerning the accuracy of AI substitution. In the context of reviewing the high ``algorithmic fidelity'' of LLMs replacing human participants, \citet{argyle2023out} explain that LLMs should require ``fewer resources than a parallel data collection with human respondents''. Similarly, \citet{wang2021want} argue that AI surrogates ``can significantly reduce labeling cost while achieving the same performance with human-labeled data''. \citet{hamalainen2023evaluating} pose this as a ``crucial question'' for user researchers: ``when can low cost and latency offset issues with quality?''}

\subsubsection{``Substitution augments demographic diversity in datasets''}

{Eight} different proposals mention that certain communities or demographic traits are of particular interest for simulation with generative AI. \citet{aher2023using} explain that substitution offers ``a simple way to simulate gender and racial diversity''. \citet{argyle2023out} contend that LLMs could generate ``patterns of attitudes and ideas'' held by ``many groups (women, men, White people, people of color, millennials, baby boomers, etc.)''{, and---along with \citet{gerosa2023can}---suggest that simulated participants can help explore intersectionality}. \citet{syntheticusers}, \citet{opinioai}, and \citet{userpersona} similarly mention synthesizing user data reflecting different genders, ethnicities, income levels, geographic locations, languages, and political views. \citet{byun2023dispensing} state simply that generative AI can be used to simulate the participation of ``vulnerable populations''. Several proposals go as far as arguing that an open-ended range of communities may be simulated, limited only by the researcher or developer’s imagination in writing a textual prompt for the LLM~\citep{syntheticusers, opinioai}.

\subsubsection{``Substitution protects participants from harms''}

{Six} proposals argue that substitutions with generative AI can protect participants from potential harm. {\citet{byun2023dispensing} summarize this idea by suggesting that simulated participants may ``prove invaluable in scenarios where human involvement would be impractical, unethical, or unsafe''. Similarly, \citet{aher2023using} discuss the use of substitution to explore research questions that would be ``unethical to run on human subjects'', offering} ``experiments on what to say to a person who is suicidal'' as an example in which LLMs could shoulder the burdens from high-risk research questions. \citet{bai2022constitutional} argue that, in AI development work, the use of LLMs reduces the need for human participants ``to engage in the rather unsavory work of trying to trick AI systems into generating harmful content''. 

\subsection{Contextualizing the motivations reported by proposals}

{To ensure a balanced evaluation of these proposals and seriously engage with the idea of substitution, we next seek to place these motivations in context. How much merit do these specific concerns carry, given the contemporary state of scientific research and AI development?}

Current funding dynamics in research and development offer validation for the first two motivations frequently cited by substitution proposals---reducing costs and increasing the efficiency of data collection. Typically, LLM users pay based on the amount of text they feed into the model and the amount of text it produces (e.g.,~\citep{openai2023pricing}). Based on standard rates, LLMs can produce data at fractions of the cost of recruiting and working with human participants. Meanwhile, financial resources are a critical constraint on social-behavioral science~\citep{almeida2023underfunding}. In HCI, for example, the need for funds to recruit and work with human participants steers a large proportion of the field into industry partnerships~\citep{critical2019patron}. Financing plays a similarly central role in the AI field. Large technology companies enjoy a great deal of influence over AI development, given the generous resources at their disposal and the smaller pools available to academic laboratories and other organizations~\citep{birhane2022values, luitse2021great, martinez2021research}. Relative affordability would thus make AI surrogates especially helpful for scientists and developers working outside of corporate laboratories.

Public dialogues around AI and science offer validation for the other two stated goals that we find manifest in substitution proposals: protecting participants from harm and augmenting demographic diversity in datasets. In traditional ethics frameworks, the principles of non-maleficence (often summarized as ``do no harm'') and justice (which seeks to ensure a fair and equitable distribution of risks and benefits from research) are paramount~\citep{apa2016ethical, national1978belmont}. These principles emerged in response to historical incidents where scientists abused their power over participants~\citep{brandt1978racism, national1978belmont}. The scientists, policymakers, and ethicists who developed these frameworks intended the concept of justice in particular to address the vulnerability of marginalized communities in the research process. With the advent of big data, interpretations of justice now also focus on ensuring that marginalized groups are (safely) reflected in research datasets and outcomes~\citep{metcalf2016human}. Non-maleficence and justice are similarly two of the foremost concerns of policymakers and ethicists over AI development~\citep{jobin2019global}. Thus, arguments to use substitution to protect participants and to improve demographic diversity in data collection echo widely shared ethical concerns over research and development.

\section{Evaluating challenges to the replacement of human participants}

{Of course, the role of a critical scholar is not to simply accept prevailing narratives at face value, but rather to interrogate them. We next turn a critical gaze to the manifest motivations offered by substitution proposals. To rigorously analyze these arguments, we first consider the technical properties and limitations of current language models. We then explore the social structure and epistemic foundations of research and development to help deconstruct the claims, promises, and potential pitfalls embedded within these proposals.}

\subsection{Practical obstacles to the replacement of human participants} \label{sec:practical_obstacles}

The substitution proposals themselves raise our first practical issue with substitution: modern language models are not yet ready to simulate human cognition and decision making. Despite their general fluency with language, LLMs routinely make mistakes across all domains to which they are applied (e.g.,~\citep{anthropic2023model, openai2023gpt4system}). Their strong tendency to ``hallucinate'' false information is a particular problem, especially because they do not indicate when they produce or cite false information~\citep{ji2023survey}. In early 2023, for instance, a U.S. lawyer asked an LLM to act as a legal assistant and subsequently submitted a court filing riddled with fabricated legal references---partially because he believed the model’s statements assuring him that the cases were real~\citep{weiser2023here}.
{Indeed, multiple proposals in our survey acknowledged the risk that hallucination introduces when simulating participants with an LLM~\citep{argyle2023out, byun2023dispensing, chiang2023can, dillion2023can, heyman2023impact, horton2023large}. This high risk of generating ungrounded responses undermines the cost and efficiency arguments for substitution, given their assumption of high accuracy levels.}

A second technical obstacle is ``value lock-in''~\citep{weidinger2022taxonomy}. The social and cultural conditions of a time period constrain and influence the ways in which people express attitudes, beliefs, and behavior~\citep{cooper1998theories}. Conceptually, LLMs develop their representations of human behavior and decision making from the norms and attitudes present in human writing (e.g., implicitly learning from the attitudes expressed on internet websites and in books). However, an LLM is typically trained just once. When the underlying attitudes and conventions shift, an LLM does not update itself to encode new norms: the responses it generates will still reflect attitudes from the time when it was trained. {An LLM trained on data before 2018, for instance, would not predict the shift in privacy concerns among social media users following March of that year~\citep{gonzalez2022regional, jones2018facebook}. Much like the constraints imposed by hallucination, value lock-in challenges the accuracy of substitution, and therefore weakens the arguments focused on efficiency and financial benefits.}

Third, modern LLMs struggle to model the wide range of opinions held across human communities, especially minority perspectives. Existing language models most readily simulate opinions from liberal, highly educated, and high-income earning individuals, and struggle to generate opinions reflective of demographic groups including individuals who are older individuals, religious, or widowed~\citep{santurkar2023whose}. As \citet{crockett2023should} point out, the training data used for LLMs echoes and reinforces the focus of psychology and HCI research on western, educated, industrialized, rich, and democratic people~\citep{henrich2010most, linxen2021weird}. Indeed, current LLMs exhibit substantial cultural bias~\citep{atari2023humans, durmus2023towards, naous2023having} and routinely produce stereotyped images of minority individuals~\citep{abid2021persistent, ousidhoum2021probing}. These myopias severely limit the suggestion that substitution can augment research and development work by simulating perspectives from marginalized communities.

Finally, user and psychology research rely on a variety of non-linguistic indicators to study and understand human cognition and behavior. The measurement of reaction time, facial expressions, and even pupil dilation have been integral in the history of psychology research~\citep{ekman1993facial, hess1964pupil, robinson2001reaction}. While it is easy to see how the text produced by an LLM can simulate participants’ text responses or multiple-choice selections, language is not a natural analog for many behavioral measures. {Though it may accelerate the speed of data collection and alleviate funding issues, substitution also restricts the types of questions available for exploration, and thus on the possible paths that research can take.}

\subsection{Intrinsic challenges to the replacement of human participants}

These four practical obstacles impose substantial constraints on the sorts of research questions and insights that LLM substitutes can support. Nonetheless, they are not inherent flaws in generative modeling. Future iterations and improvements to models could discover novel solutions to address hallucinations, the preclusion of minority perspectives, and value lock-in. Even the problem of non-linguistic measures may admit an eventual solution through clever uses of language (e.g.,~\citep{yao2022react}) or multimodal generative AI (e.g.,~\citep{ho2022imagen}). Technical progress may alleviate these concerns, and perhaps overcome them entirely.\footnote{Indeed, modern AI research has frequently defied expectations, sometimes on remarkable timescales. For example, see the survey conducted by~\citet{grace2018will}, collecting predictions of important milestones in AI development from AI scientists. Multiple breakthroughs handily subverted their corresponding expert-predicted timelines.} {In addition, while these obstacles undermine three of the four motivations revealed in our review, they do not erode the argument for protecting participants from risk. A researcher interested in experimenting with AI surrogates might consider such protections worth the costs.}

However, practical issues are only part of the picture. We next critically appraise the broader context of human participation. This exploration reveals tensions that are considerably more fundamental in nature. In particular, substitution proposals conflict with values intrinsic to AI development work and research science: namely, representation, inclusion, and understanding.

\subsubsection{The values of representation and inclusion}

First, replacing human participants with LLMs inherently conflicts with \textit{representation} and \textit{inclusion}, two principles core to the participatory development of AI technology.

Discussing the values of participation, representation, and inclusion can be difficult, given both the ambiguous definitions of the underlying concepts and the way that researchers and developers often use the terms interchangeably~\citep{birhane2022power, chasalow2021representativeness, cornwall2008unpacking}. More importantly, invocations of all three concepts have been criticized for connoting the involvement or presence of people without any substantive changes over the status quo~\citep{barton2020beyond, bergman2023representation, birhane2022power, chasalow2021representativeness, cornwall2008unpacking, quick2011distinguishing}. To paraphrase \citet{cornwall2008unpacking}, being involved in a process is not equivalent to having any voice or any power over it. Participation in particular is often defined shallowly, with ``participant'' reflecting an incidental sort of role, resulting simply from being part of the environment or process that leads to a technological product~\citep{groves2023going, jing2023towards}. Whether intentional or not, shallow uses of the ``participatory'' label produce a veneer of representation and inclusion, despite the lack of deeper engagement with participating individuals and communities.

Here we focus on ``participation'' not in these shallow forms, but as the concept was originally conceived in public planning and development. In a foundational treatise, \citet{arnstein1969ladder} articulated a vision of participation that is ``responsive to'' and reflective of the ``views, aspirations, and needs'' of those participating. For Arnstein, authentic participation must involve ``the redistribution of power that enables the [have-nots], presently excluded from the political and economic processes, to be deliberately included in the future''. This description reveals two intertwined goals for participatory processes: first, to reflect and respond to the interests of those participating; and second, to redistribute power to those participating. We refer to these attributes as representation and inclusion, respectively, based both on Arnstein’s arguments\footnote{Arnstein’s treatise is foundational in participatory design, but did not introduce these ideas for the first time. The intellectual lineage of the ideas of responsiveness and power traces back several decades before Arnstein’s piece, emerging across multiple schools of thought in the history of planning~\citep{lane2005public}.} and on established frameworks from political science and organizational theory. The goal of representation involves an active process of understanding and ``making present'' the needs, interests, and perspectives of those being represented.\footnote{In particular, \citet{pitkin1967concept} describes representation as "acting in the interest of the represented, in a manner responsive to them''. \citet{arnesen2018legitimacy} offer an eloquent elaboration of this idea, defining ``substantive representation'' as the situation where the ``‘making present’ of [a] group’s ideas and the ‘acting for’ that group is done by people who understand what it means to be part of this group''.} In contrast, the goal of inclusion emphasizes participants’ agency and exercise of power.\footnote{Concerning inclusion, \citet{acemouglu2016paths} argue that inclusive institutions distribute power broadly, empowering a wide range of people to participate in their activities. Similarly, \citet{dahl2008polyarchy} recognizes inclusiveness as a fundamental dimension of democracy, defining it as the participation of all members of society in the exercise of power. In organization theory, \citet{quick2011distinguishing} muse that ``inclusion practices'' must remake rather than recreate existing power structures in order to truly invite communities to ``co-produc[e] processes, policies, and programs''.} Synthesizing these aims, participatory work should draw in participants to collaboratively set the agenda, empower them with influence over the development process, and guarantee them avenues of recourse, reparation, and resistance.

Corroborating this conceptual framework, prominent AI institutions frequently advocate for participatory approaches to development that explicitly tie participation to the goals of representation and inclusion. \citet{zaremba2023democratic}, for instance, argue that future AI systems should ``be shaped to be as inclusive as possible'' through democratic and representative processes. Other organizations echo similar plans, including one advocating for ``democratic participation'' to develop principles to govern the behavior of future AI systems~\citep{roose2023dario}. \citet{glaese2022improving} likewise appeal for ``participatory input from a diverse array of users and affected groups'' for the development of more advanced systems. These appeals signify that the participation of individuals and communities is not just about improving data collection, but also about representation and empowerment. Turning these ideas back to substitution proposals: simulating human participants with LLMs may support more efficient data collection, but does it embody these participatory goals?

Unfortunately, substitution proposals only support representation in a very weak sense. At some point in their training, the models process linguistic data from humans, and development teams subsequently use these models to stand in for the perspectives and opinions that helped produce those data. But how well do the modeling outputs reflect people’s perspectives, and how responsive is the substitution process to changes in those views? The current class of substitution proposals do not specify any possible approaches to re-align a wayward LLM, and previous attempts to manually steer models observed limited improvements to representativeness~\citep{santurkar2023whose}. Without mechanisms to address these questions, substitution is not truly representative---it does not ``make present'' people’s experiences.

As for inclusion: if the sharing of power is core to inclusiveness, then the current class of substitution proposals are inherently exclusionary. Participation in the development process, even through data enrichment, provides people with critical capabilities and affordances. These include the right to withhold data, the right to opt out, the right to express discontent, and the right to resist, to name a few~\citep{agnew2023technologies}. Substitution proposals shift several of these powers to AI surrogates, thereby denying participants agency over mechanisms of feedback and change---and implicitly ceding control to the developers. Some rights are lost altogether. When faced with an unethical situation, for instance, LLMs will not report the issue via regulatory authorities, media organizations, or other institutions that could bring the situation to light.

These exclusionary dynamics are especially important given that several proposals recommend simulating racial minorities, women, and other marginalized communities to help address issues with data diversity~\citep{argyle2023out, byun2023dispensing}. These communities have long, painful histories of exclusion and exploitation in scientific and technological development~\citep{national1978belmont, scharff2010more}.
{To ignore this historical context is to risk conflating technological change with social progress, devoid of any ``moral and political standards''~\citep{marx2010technology}. Indeed, scholars in the humanities argue that visions of replacing human labor with AI perpetuate troubling social and political legacies. The types of labor considered ``appropriate'' for automation are typically those already devalued and exploited under capitalism~\citep{rhee2018robotic}. As a consequence, the idea that new technology will render human labor obsolete echoes colonial fantasies of an ideal servant class, working invisibly without need or desire~\citep{atanasoski2019surrogate}. Ironically, the very conceit of simulating marginalized communities as a solution to exclusion reinforces the devaluation of their labor and lived experiences.}

Ultimately, replacing participants denies them opportunities to {directly combat this devaluation,} to influence the goals of the development process, {and to help adjudicate the balance between their rights and the risks that they experience}.
It may be the case that the individuals and communities involved in development work wish to relinquish their participatory role. Rather than entirely turning to technological solutions, a truly representative and inclusive process would explicitly ask that question of them. %

\subsubsection{The value of understanding}

The second intrinsic challenge to substitution proposals concerns the foundations of user research and psychological science. Replacing human participants with AI disrupts the intersubjectivity between researcher and participant, undermining the goal of \textit{understanding}.

A core aim of research science is the production of generalizable insights and knowledge (cf.~\citep{mook1983defense}). Indeed, in the U.S., government regulation highlights the goal of ``develop[ing] or contribut[ing] to generalizable knowledge'' as a definitional characteristic of research~\citep{us2018code}. Scientific culture often idealizes research as an objective process that leverages precise measurement and logical interpretation to uncover general insights about the natural world~\citep{reiss2014scientific}. From this perspective, psychological fields observe and measure external indicators like behavior to better understand internal processes like cognition and decision making.

In reality, the basis of psychological research and insight is not objective measurement, but intersubjective corroboration~\citep{mascolo2020phenomenology}.
Consider a user experience (UX) researcher in their laboratory, conducting a study on user preferences over several possible website designs. 
Each user in the laboratory’s study room sits in front of a computer and sees two designs in sequence. They then respond to a question asking for their relative preference between the two on a scale from one to seven. In an objectivist account, the user’s preferences are a construct that exists in the real world, but which the researcher cannot directly access. By measuring an indicator that is externally observable and quantifiable, the researcher can estimate the user’s preferences (and then test various hypotheses over that construct).

But how does the researcher know the meaning of the external indicator in the first place? If internal states and constructs are truly private, why does the researcher believe that acting in a certain way (e.g., interacting with the mouse and selecting one of the numbers on the scale) maps to some specific internal experience (e.g., a preference for one design over the other)? The answer is that the researcher does not deduce from external measurements alone. Rather, they combine those ``objective'' indicators with their own internal experience to draw an inference about the participant.
The idea of intersubjectivity can seem abstruse on first encounter, but it can be summarized simply as the researcher’s assumption that if they were to take another person’s position, they would see or experience the situation as that person does. In short, intersubjectivity is ``the possibility of trading places''~\citep{duranti2010husserl}.

Sometimes, it is more or less credible for two individuals to ``trade places''. Variations in this possibility---that is, in the degree of intersubjectivity---constrain and shape a scientist’s ability to conduct research.\footnote{Indeed, it is for this reason that philosophers of science sometimes distinguish the social sciences as different in kind from the natural and physical sciences~\citep{mayntz2005forschungsmethoden}. Unlike psychologists and other social scientists, natural and physical scientists have no intersubjective relationship with the objects of their study. A physicist cannot trade places with a particle, and a geologist cannot trade places with a rock. These scientists cannot (and need not) corroborate their knowledge claims with intersubjectivity. While all scientists seek to develop knowledge of the world, the study of humans demands a special kind of knowledge: \textit{Verstehen}, or a ``participatory'' type of understanding~\citep{weber1947theory}.} For instance,
{an HCI researcher} traveling and encountering a foreign culture for the first time would find it very difficult to draw sound inferences about user behavior from its inhabitants without first spending more time studying local norms and ways of living (e.g., {\citep{bourges1998meaning}}). Situations with limited intersubjectivity require researchers to continuously question and challenge their own assumptions.

To return to the topic of AI substitution, let us revisit our hypothetical UX laboratory. %
Our intrepid researcher remains in the laboratory late one evening to analyze their data. As they diligently work, an alien appears in the study room in front of the computer. It moves its arms---or what \textit{might} be its arms---and in doing so, nudges the mouse, pressing down on its buttons several times. These actions cause the computer to flip through both websites before registering a two on the researcher’s seven-point preference scale. Can our researcher thank the alien, not just for the fame of first contact, but also for the free data provided to their study? Should they add a tally to the count of participants who prefer the first design? Perhaps the alien does prefer the first design, much like the human participant who visited earlier. However, the crucial intersubjective assumption is missing. The researcher has no actual assurance that internal preferences drove the alien’s action, or indeed that the alien possesses something resembling human preferences. Until our researcher gains a much deeper understanding of this alien, attempts to transfer any observations to human cognition are made on shaky grounds.

Language models, of course, are not alien---they are trained on human data.\footnote{Though as demonstrated by the recently popularity of the ``shoggoth'' meme~\citep{hlntnr2023you, jacyanthis2023here, tetraspacewest2022here}, many members of the AI community are skeptical of the human-likeness of LLMs.} Perhaps they share some form of an intersubjective relationship with research scientists. As in other situations with limited intersubjectivity, research with AI surrogates might then support careful insight into human cognition and decision making. For instance, {when prompted to choose from a list of possible consumer options, LLMs demonstrate a fairly reliable preference for the first option (e.g.,~\citep{kirshner2024gpt}). A researcher might infer from this model behavior---and from the general way that LLMs are trained on data from humans---that the average consumer is more likely to be swayed by the initial option presented in a user interface (cf.~\citep{mantonakis2009order}).} Ultimately, though, the credibility of trading places with language models and experiencing a situation as they do seems quite limited~\citep{bender2020climbing, shanahan2022talking}. And importantly, any intersubjective relationship that a human researcher shares with a language model is different in kind from the relationship between a human researcher and a human participant. Behavioral scientists generally maintain a good sense of the limits of intersubjectivity with their models. For example, researchers are keenly aware of the limited external validity of rule-based models (e.g.,~\citep{jackson2017agent}). However, LLMs present a distinctly powerful risk of anthropomorphism, even for researchers~\citep{weidinger2022taxonomy}. Their language proficiency can challenge and distract from the precept that ``a map is not the territory''.
Overall, the limited intersubjectivity that researchers share with generative AI systems constrains the possibility of substituting AI for human participants in research. {Improvements in efficiency and acceleration do not actually benefit research if substitution ultimately produces unsound conclusions.}

\subsubsection{Intrinsic challenges across research and development}

Thus far, we have primarily discussed the goals of representation and inclusion in the context of AI development, and the aim of understanding in the context of research science. However, these issues cross over these boundaries, emerging across both domains.

Representation and inclusion are also core to user research, psychology, and other social and behavioral fields. Participatory HCI~\citep{vines2013configuring}, computer-supported cooperative work~\citep{ciolfi2023computer}, participatory action research~\citep{fals1991action}, and critical psychology~\citep{walsh2014researcher} argue that scientists should avoid viewing participants primarily as sources of data, whose only role is to accept certain inputs and subsequently produce outputs of a desired form. Rather, they argue, researchers have ethical obligations to partner and share power with their participants. Here too, substitution proposals undermine the processes meant to make present participants' interests and erode participants' ability to influence the overall undertaking.

Similarly, intersubjectivity plays an essential role in modern AI development---particularly in its reliance on data annotation and enrichment. As \citet{lamber2023reinforcement} explain, ``The core challenge in data labeling is to make the annotators understand the task the same way you do''. At the design stage of an annotation study, in order to construct appropriate instructions, interfaces, and questions, developers must have an appreciation for their annotators’ perspectives and experiences. Similarly, at the interpretation stage, developers must understand their annotators’ perspective to make sense of their results and identify any issues that could interfere with their ability to interpret their annotations. Data annotation and enrichment are predicated on the possibility of ``trading places''.

\subsubsection{Summary}

Overall, the values of representation, inclusion, and understanding pose considerable challenges to substitution proposals. These challenges are different in kind from the initial obstacles discussed in Section~\ref{sec:practical_obstacles}. Conflicts with representation, inclusion, and understanding cannot be alleviated with better training or improved model performance alone: they are intrinsic to the models and to the proposed approach to replacement itself.

\section{Conclusion and paths forward}

Large language models (LLMs) and other generative artificial intelligence (AI) systems will likely change the way that people approach many professional tasks. Research science and technology development are no exception. Recently, a set of curious proposals has emerged, recommending the substitution of LLMs for human participants in AI development and in research contexts ranging from human-computer interaction to opinion polling. These proposals carry several potential merits: increasing the speed of research and development; reducing financial costs; mitigating participants' exposure to potential harms; and augmenting demographic diversity. However, {three of} these merits are constrained by technical shortcomings of current AI models. More pressingly, replacement faces two intrinsic challenges from the broader social and epistemological structures that support human participation. First, while participatory development aims to ``make present'' participants’ perspectives and share power with them, substitution proposals have the opposite effect---impeding representation and inclusion. Second, substitution proposals remove the intersubjective basis of psychological and user research, limiting the inferences that psychologists and researchers can draw from studies and experiments with LLMs. These {undercut all four stated motivations for substitution, and are thus} considerable challenges to any proposal for the replacement of human participants in AI development and psychology research.

A deeper reckoning with the values of representation, inclusion, and understanding is needed to identify changes that could mitigate these intrinsic challenges. A first step toward representative and inclusive versions of substitution will be to involve and empower participants in the agenda-setting process. After formally setting the agenda, projects can maintain an ongoing role for participants, allowing them to monitor and supervise whether their substitutes are making present their perspectives and concerns. This practice of continuous engagement should strive not only to align the development process with those participating, but also to allow participants to exercise power over the development process: provide feedback, express discontent, and seek recourse. To make substitution compatible with the goal of understanding, user researchers and psychologists will need to confront a potentially uncomfortable idea: that scientists are not immune to the temptation to anthropomorphize language models. One potential approach to mitigating the risk of anthropomorphism would be to establish a clear framework for evaluating the results of substitution experiments and assessing whether they generate any valid insights. Such a framework could help researchers remain mindful of the limits on ``trading places'' with language models. Overall, by establishing these structures and practices, we may start to build generative AI systems that support the values fundamental to research and development, moving toward authentic---rather than artificial---inclusion.

\begin{acks}
We thank William Isaac, Piotr Mirowski, our anonymous reviewers, and our anonymous area chair for insightful feedback on our manuscript.
\end{acks}

\section*{Author Contributions}
W.A., J.P., and K.R.M. proposed the research idea; K.R.M. supervised the project; W.A. and K.R.M. curated and annotated sources, with assistance from A.S.B., J.C., M.D., S.E., J.P., and S.M.; W.A., A.S.B., J.C., M.D., S.E., J.P., S.M., and K.R.M. interpreted results; and W.A., A.S.B., J.C., M.D., S.E., J.P., S.M., and K.R.M. wrote the paper.

\bibliographystyle{ACM-Reference-Format}
\bibliography{main}

\appendix

\end{document}